\documentclass[12pt]{article}

\usepackage{scicite}
\usepackage{multicol}
\usepackage{subcaption}
\usepackage{empheq,etoolbox}
\usepackage[export]{adjustbox}
\usepackage{graphicx}
\usepackage{amssymb}
\usepackage{epstopdf}
\usepackage{hyperref}
\usepackage{ulem}
\usepackage{amsmath}
\usepackage{caption}
\usepackage{floatrow}

\usepackage{times}

\newenvironment{sciabstract}{%
\begin{quote} \bf}
{\end{quote}}

\title{Networked Infectiousness: \\Cascades, Power Laws, and Kinetics}

\author
{Sara Najem$^{1,2\ast}$, Leonid Klushin $^{1,3}$, and Jihad Touma$^{1,2\ast}$\\
\\
\normalsize{$^{1}$Department of Physics, American University of Beirut, Beirut, Lebanon}\\
\normalsize{$^{2}$Center for Advanced Mathematical Sciences, American University of Beirut, Beirut, Lebanon}\\
\normalsize{$^{3}$ Institute of Macromolecular Compounds RAS, St.Petersburg, Russia}\\
\\
\normalsize{$^\ast$To whom correspondence should be addressed; E-mail:  sn62@aub.edu.lb, jt00@aub.edu.lb.}
}

\date{}

\begin{document} 


\baselineskip24pt


\maketitle


\begin{sciabstract}

Networked SIR models have become essential workhorses in the modeling of epidemics, their inception, propagation and control. Here, and building on this venerable tradition, we report on the emergence of a remarkable self-organization of infectiousness in the wake of a propagating disease front. It manifests as a cascading power-law distribution of disease strength in networked SIR simulations, and is then confirmed with suitably defined kinetics, then stochastic modeling of surveillance data. Given the success of the networked SIR models which brought it to light, we expect this scale-invariant feature to be of universal significance, characterizing the evolution of disease within and across transportation networks, informing the design of control strategies, and providing a litmus test for the soundness of disease propagation models.

\end{sciabstract}

% In setting up this template for *Science* papers, we've used both
% the \section* command and the \paragraph* command for topical
% divisions.  Which you use will of course depend on the type of paper
% you're writing.  Review Articles tend to have displayed headings, for
% which \section* is more appropriate; Research Articles, when they have
% formal topical divisions at all, tend to signal them with bold text
% that runs into the paragraph, for which \paragraph* is the right
% choice.  Either way, use the asterisk (*) modifier, as shown, to

Think epidemic and you will find yourself asking: when will it reach me and where did it originate, and eventually what measures should one take to mitigate it? These questions are as old as disease spreading from villages, to towns, and wider through travel, trade, pilgrimage, war…etc. There are to be sure critical timescales at play: some disease-specific and others associated with the mode of circulation. And with circulation one has to further worry about the topology of the transportation network, in addition to traffic flow through it\cite{newman2002spread,serrano2006percolation,pastor2015epidemic}. While disease-specific timescales and virulence have always shown significant variability, it is the scale, topological complexity and efficiency of transportation networks that have accelerated dramatically in recent times connecting the planet to itself over timescales of hours, to replace the years of times past \cite{bhattacharya2019social,brockmann2006scaling,gonzalez2008understanding,castellano2009statistical,simini2012universal,zarocostas2020fight,barnes2014big}. Couple disease spread to information spread at the speed of light, and you have the makings of global lockdown witnessed in the recent COVID pandemic \cite{balakrishnan2022infodemic,rovetta2022we,atehortua2021covid}. In the face of such overwhelming complexity, it is of course natural to resort to large scale simulations which integrate surveillance data to allow prediction and guide mitigation efforts, all the while informing fundamental understanding of that same complexity \cite{broeck2011gleamviz,vespignani2009predicting,balcan2009multiscale}. In dialogue with this fine-grained simulation landscape, the seminal work of Brockmann and Helbing (BH for short) reshaped our understanding of modes of disease propagation, with profound implications for modeling of epidemics, numerical and otherwise \cite{brockmann2013hidden,manitz2014origin}. Their insightful mapping of geographic distance into a traffic-weighed effective distance would morph the apparent pandemonium of disease spread over a complex network into the appealing picture of a near-circular reaction-diffusion front spreading through a flammable infection-prone medium. Thus, if worried about disease arrival time, follow the front; and if concerned about disease origin, locate the front's center and you shall have it! There followed a series of natural variations within the BH framework, then others allowing for control measures over networks which result in time-evolving topology then heterogeneity of time of arrival \cite{selley2015dynamic,pare2017epidemic,nadini2018epidemic,holme2014birth,masuda2013predicting,sayama2013modeling,valdano2015analytical}. 

In this work, we were inspired by remarkable features of surveillance data \cite{najem2022framework}, then the elegance of the BH framework to go beyond legitimate (though excessively reductionist) concerns with the origin of an epidemic and the time of its arrival. Instead, we focus on the complement, the bulk of its evolutionary dynamics, namely the self-organization of infectiousness in the wake of the reaction-diffusion front as it cascades over the network. 

%And here, we were pleasantly surprised to witness the emergence of a power law distribution which is characteristic of such cascades, then went on to construct a kinetic theory for its emergence. The whole was then confronted with stochastic modeling of COVID surveillance data with sufficient agreement to confidently announce the identification of a universal regime of disease propagation. But first things first, BH for a start, then revelations to follow. 

With BH, local disease dynamics are captured with a vanilla-type SIR model, and coupling between nodes is assured via the transport of infected individuals across the network: 
\begin{eqnarray}
\partial_{t}j_{n} & = &a v_{n}j_{n}-b j_{n}+c\sum_{m\neq n}P_{mn}\left(j_{m}-j_{n}\right),\nonumber \\
\partial_{t}v_{n} & = &-a v_{n}j_{n}+c \sum_{m\neq n}P_{mn}\left(v_{m}-v_{n}\right),
\label{eq:model}
\end{eqnarray}
where $v_{n}$, $j_{n}$, and $r_{n}$  are the fractions of susceptible (vulnerable), infected, and recovered individuals in a locality $n$, with $r_{n}=1-v_{n}-j_{n}$; $a$ and $b$ are the contact and mean recovery rates, respectively. Traffic $P_{mn}$ between nodes $m$ and $n$ follows the gravity model $P_{mn} = I_m F_n D_{euc,mn}^{-2.6}/\sum P_{mn}$, and is a function of the Euclidean distance between nodes, $D_{euc,mn}$ ,  the populations $I_m$ and $F_n$ at $m$ and $n$ respectively. This choice is substantiated by data-driven models of human activity \cite{manitz2014origin,meyer2014power,zipf1946p,simini2012universal,barthelemy2011spatial}.% and the average mobility rates

As noted above, BH replace geographical distance by an effective pairwise distance $d_{mn} = (1 - \log{P_{mn}})$ in which geographically distant nodes $m$ and $n$ are effectively close if the traffic $P_{mn}$ between them is heavy and, conversely, geographically close nodes are effectively distant if the traffic between them is light. One can then define a directed path length $\lambda(\Gamma,P)$ as the sum of the lengths along a given path $\Gamma$, an ordered set of nodes, which leads to the definition of the shortest effective path, given the effective distance $d_{mn}(P_{mn})$, as $D_{eff} = \underset{\Gamma}{\min} \ \lambda(\Gamma,P)$.   This remapping of geography into effective distance restores order to the seemingly random flow over geography by revealing a (reaction-diffusion) wavefront which is centered at the source node, and propagates in nearly concentric circles along the shortest paths to subsequent nodes. Those circles are not detectable over geography nor are they from the perspective of any node $m \neq n_{o}$, where node $n_{o}$ is the origin of the outbreak. In addition, arrival times $T_a(m|n)$ were shown to correlate linearly with $D_{eff}$, allowing the recovery of a speed of propagation given by the ratio $v_{eff} =  D_{eff}(P)/T_a$. While $D_{eff}$ is governed by topological properties of the static traffic network $P$, $v_{eff}$ is a function of global epidemiological parameters, and was shown to be independent of $P$ and of the origin of the outbreak.

%In sum, BH were able to reliably predict arrival times, then locate the source of an outbreak by reversing the arrow of time. 

%Thus, the latter corresponds to the node in which the correlation between $T_a$ and $D_{eff}$ is maximal or equivalently the average distance between the potential outbreak location and the other nodes and their standard deviations are simultaneously minimized.

\begin{figure}[!htp]
 % \begin{minipage}{\textwidth}
 \centering
\begin{subfigure}[b]{0.475\textwidth}
        %  \centering
        \includegraphics[scale=0.42]{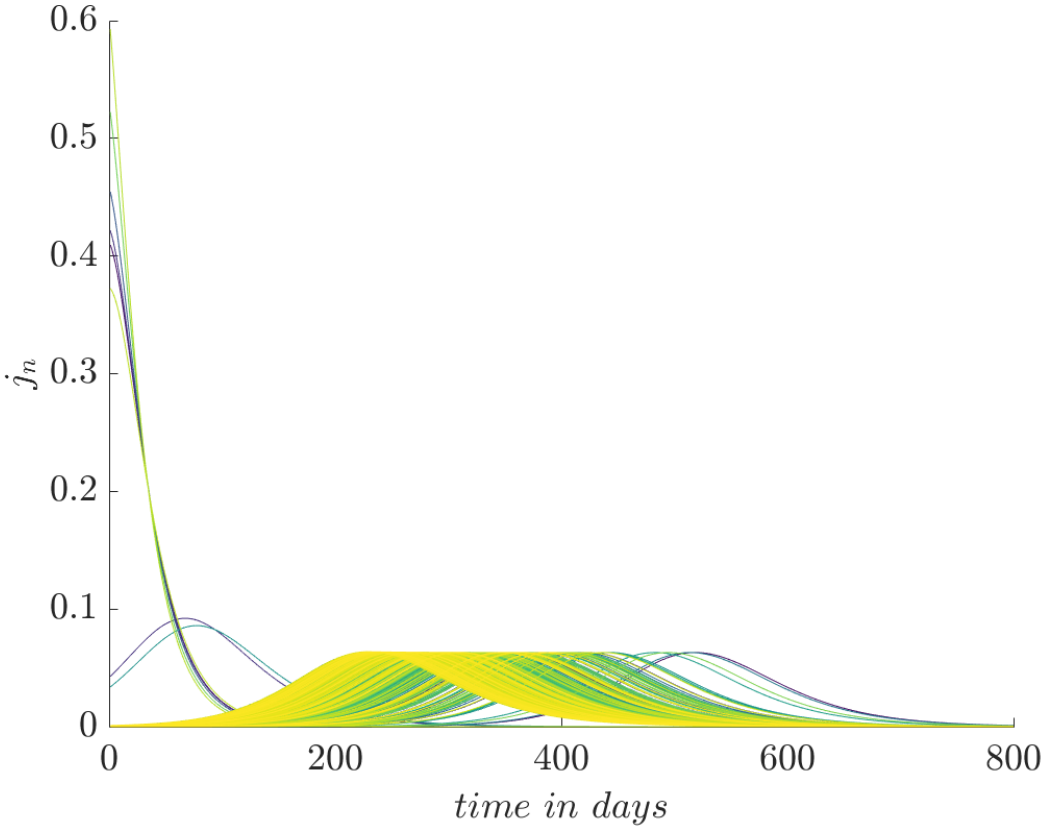}%JnNoSourceOriginKnown2Light.pdf
        \caption{}
         \label{bhjnt}
     \end{subfigure}
    %  \hfill
          \begin{subfigure}[b]{0.475\textwidth}
        %  \centering
    \includegraphics[width=0.8\textwidth]{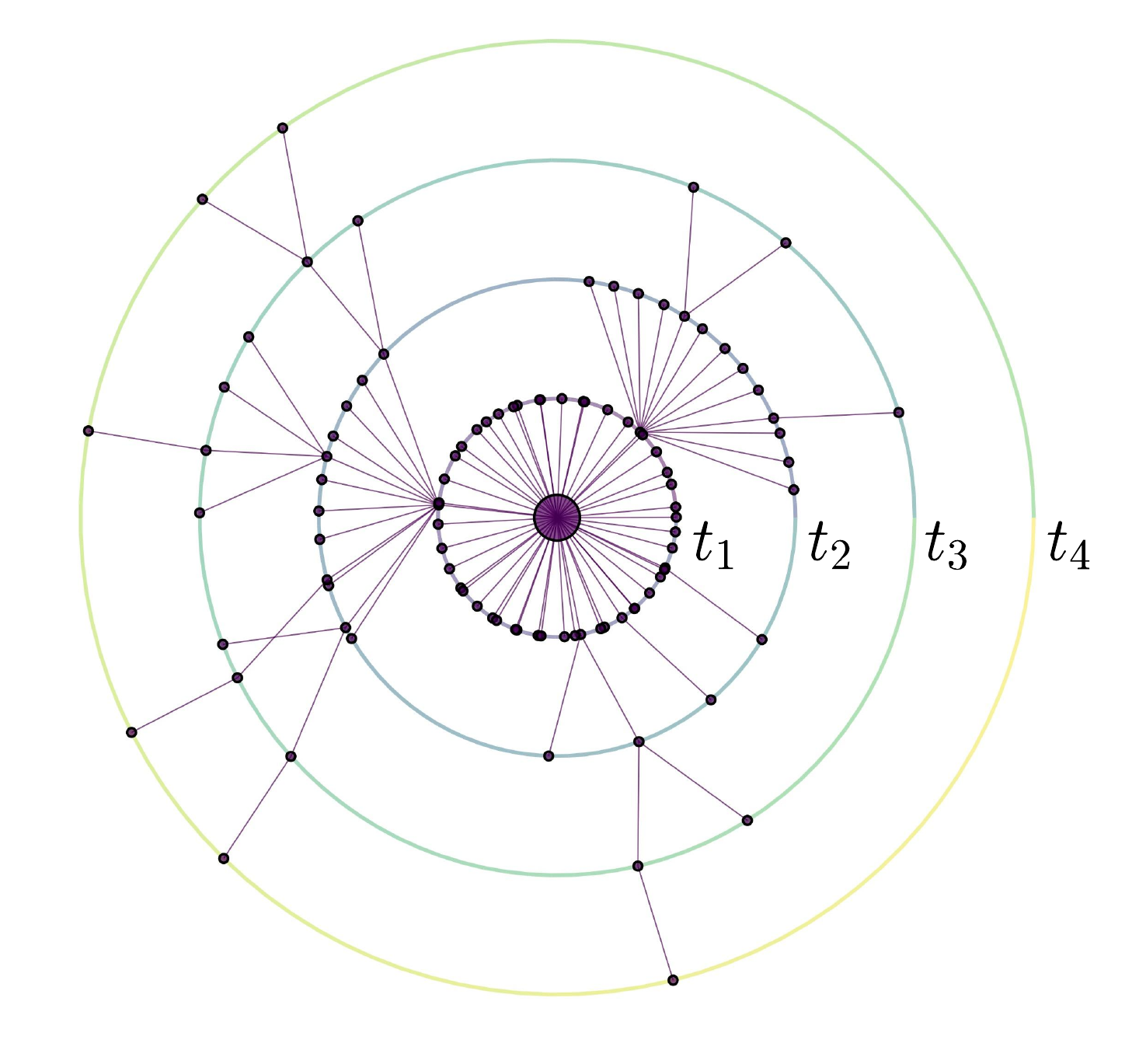}
    \caption{}
    \label{fig:wavecon}
     \end{subfigure}
     %140,175,225,275
\caption{ Eq.\ref{eq:model} is solved over Lebanon, which has a network of size  $N=1544$  with the following choice of parameters  $b= 0.05$, $a= 1.5b$, $c = 1e^{-6}$. Fig. \ref{bhjnt} follows the time evolution of $j_m(t)$.  Using {\sf  NetOrigin}, an {\sf R} package which implements the BH framework, the capital Beirut [node $n_o = 410$ in our network] is identified as the origin of the epidemic, and disease propagation shown to be structured along circular fronts, with nodes at equal effective distance, $D_{eff}$, from $n_o$. Nodes and fronts are shown in 
\ref{fig:wavecon} at $t \in \{t_1=280,t_2 = 320,t_3= 360,t_4=400\}$, and respective at $D_{eff} = \{10.96, 11.10, 11.31, 11.39\}$ from the source node.}  %The total infection counts at the fronts are $\{10897,3707,1936 ,990\}$ sequentially.}
\label{fig:typical}       % Give a unique label
\end{figure}

 In what follows, we simulate Eqs.\ref{eq:model} for Lebanon with a network of size $N = 1554$ consisting of the smallest administrative units in the country. To determine $P_{mn}$ between localities, we estimate Euclidean pairwise distances and use the populations $I_m$ and $F_n$ as in \cite{najem2022framework}. Localities in the network are seeded with infections which reflect their prominence in the traffic network, and the dynamics allowed to play out {\bf[SM, Sec. \ref{sec1app}]}. Two distinct perspectives on disease propagation are presented in Fig. \ref{fig:typical}.  In Fig. \ref{bhjnt}, we follow the evolution of infection count $j_m(t)$ over the network. One notes three distinct phases in the evolution: i- an initial phase, around $[0, 250]$ days, with infection and recovery of initially infected hubs in the network, accompanied with the more gradual growth of infection in remote nodes via transport over the network. ii- An intermediate phase which kicks in at the tail end of those initial peaks, then features a distinctive plateau in infectivity around $[250,550]$ days. This plateau reflects a node-independent, pattern of growth and decay in $j(t)$ with increasingly distant nodes taking turn hitting the same maximal infectivity, a scale invariant behavior in what appears to be a fine balance between infection recovery and in-out transport. We note that, nearly 400 days after its inception, the epidemic permeates the whole network, with the farthest nodes taking till $t \approx 550$ days to reach their maximum infection count; iii- a final stage, with all nodes in the network having peaked in infectivity, and now proceeding to full recovery. Feeding simulation results into an open-source implementation of BH machinery [namely the {\sf  NetOrigin} package\cite{manitz2018netorigin}], we were not surprised to learn that the capital Beirut (node $n_o=410$) is identified as the source of the epidemic in our fiducial experiment, given that it is the most populous and densely connected node in the network.  The same machinery yields the location of effective wave fronts, concentric circles centered at $n_o=410$,  which we show for four distinct arrival times in Fig. \ref{fig:wavecon}, the last being $t_4=400$ days around which the epidemic has covered the whole network. We then situate those fronts in relation to the underlying network, by showing a sample of nodes partaking of a given front, and the edges connecting them. Keep in mind that nodes on fronts are at the same effective distance from the node of origin, with nodes reached by $t_4$ located around the maximum effective distance from Beirut, $D_{eff} \sim 11.4$.

The analysis thus far has focused on classical measures and yielded: i- the recovery of the origin of the epidemic, then arrival times along circular fronts within the BH framework which has been largely integrated in epidemiological studies; ii- and, perhaps more interestingly, an emergent, scale invariant pattern of growth and decay of infection counts across the transportation network which to our knowledge, has not been reported in the literature to date. Interestingly enough, the inception of the $j$-plateau coincides with the emergence of BH fronts around $t = 280 $ days, with its range extending beyond disease permeation of the network around $t = 400 $ days. We relegate full investigation of this curious regime to a separate publication, and instead focus on yet another global feature of networked disease propagation, and the kinetic theory that captures it, highlighting further correlations with BH-front and $j$-plateau along the way. 

\begin{figure}[!htp]
  % \begin{minipage}{\textwidth}
  \centering
     \includegraphics[scale=0.04]{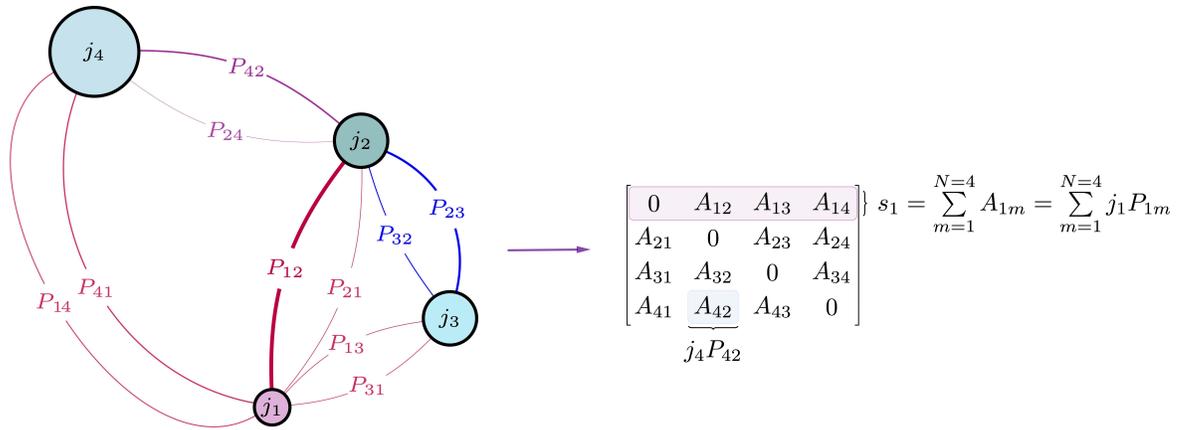}%\quad  \hfill \hspace{-1.2cm}
   \caption{We show a sub-graph of 4 localities on the traffic network, and its corresponding matrix representation. Nodes on the graph are characterized by the fraction of infected population $j_n$, and edges by traffic intensity $P_{nm}$, where $n$ and $m\in \{1,2,3,4\}$. The matrix element, highlighted in blue, captures the flux of infected individuals between its corresponding row and column localities. The sum of the columns for a given row, or locality, highlighted in light purple, is the node's strength and corresponds to the net flux of infected individuals from that node.}  \label{definition}
\end{figure}

Starting with BH dynamics over a static transportation network, we construct a time-varying network which encodes invaluable information on disease kinetics. The various layers of our constructions are illustrated in Fig. \ref{definition}. We first build the matrix $A(t)$ with elements $A_{nm} (t) =  j_n(t) P_{nm}$, coupling the time-varying infection at a given node $n$ to the transport of infection from that specific node to any other node $m$\cite{xiong2020mobile,sulyok2020community,iacus2020human}. We then recover the sum of infection-weighed links emerging from $n$ in the traffic network, $s_n(t) =  \sum \limits_{m} j_n(t) P_{nm}$, which is the strength of node $n$ and effectively its infectivity export \cite{barrat2004architecture}.  
For us, the degree distribution of the (static) traffic network, an essential topological determinant in outbreaks, will be replaced by the distribution of node strengths, $s_n(t)$, in $A(t)$. In fact, a key result of our work pertains to the remarkable self-similar nature of this distribution as the disease cascades over the network. What is our evidence for this self-similarity? and what is the disease-driven kinetics that shapes it?

\begin{figure}[htbp]
\centering
\vspace{-4cm}
% Left minipage with single centered figure
\begin{subfigure}{0.5\textwidth}
\vspace{4cm}
\begin{minipage}[t]{0.52\textwidth}
    % \centering
\includegraphics[width=1.8\linewidth,keepaspectratio]{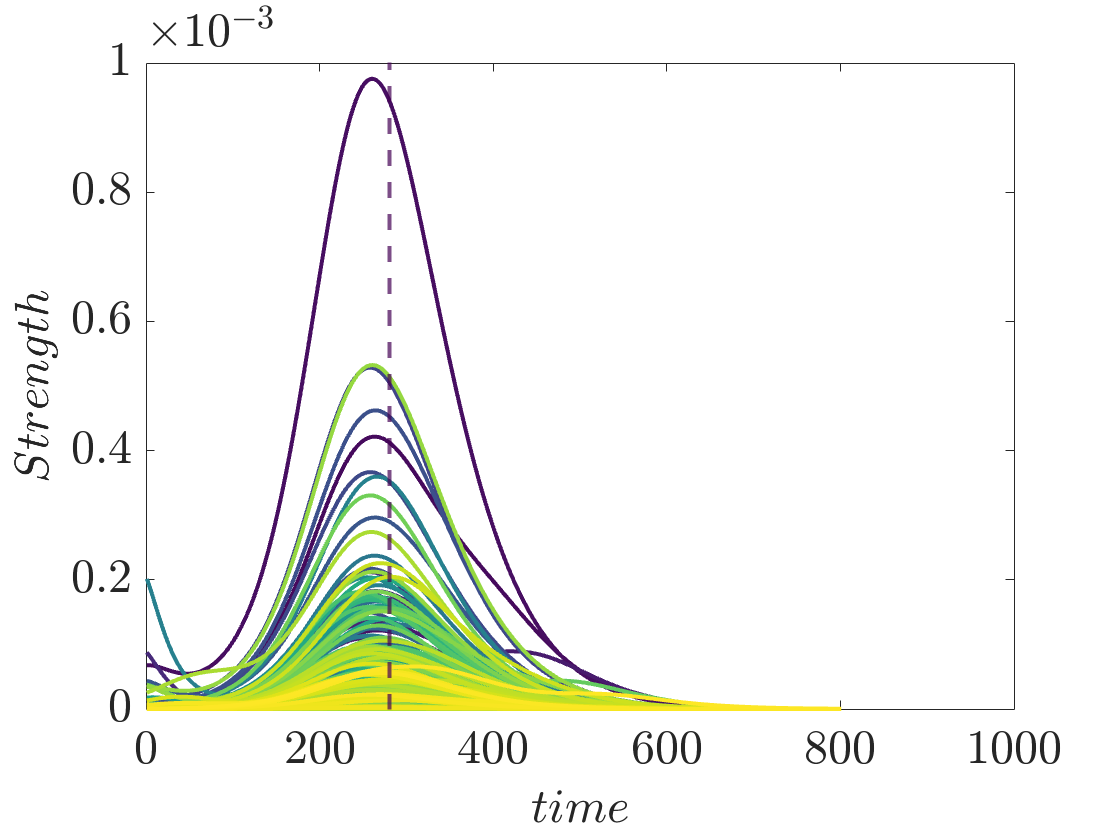}
\caption{}
    \label{fig:left}  
\end{minipage}
  
\end{subfigure}
\hfill
% Right minipage with subfigures
\begin{minipage}[t]{0.42\textwidth}
    % Top row: 3 subfigures
    \begin{subfigure}{\textwidth}
        \centering
        \begin{minipage}[t]{0.42\textwidth}
            \centering
            \includegraphics[width=\linewidth,height=3.5cm,keepaspectratio]{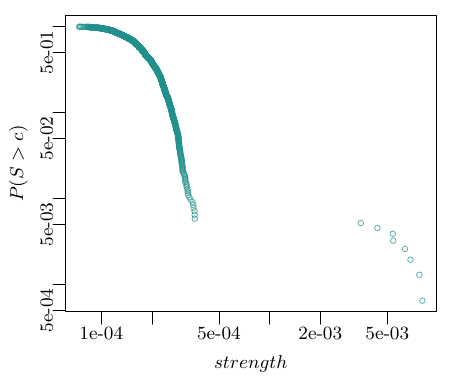}
            \includegraphics[width=\linewidth,height=3.5cm,keepaspectratio]{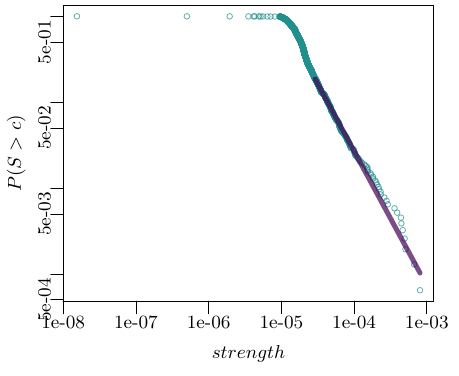}

        \end{minipage}
        % \hfill
        % \begin{minipage}[t]{0.32\textwidth}
        %     \centering
        %     \includegraphics[width=\linewidth,height=3.5cm,keepaspectratio]{strengthtim28hub.pdf}
        %     % \caption{Data preprocessing pipeline stages}
        %     \label{fig:dataflow}
        % \end{minipage}
        \hfil
        \begin{minipage}[t]{0.42\textwidth}
            \centering
            \includegraphics[width=\linewidth,height=3.5cm,keepaspectratio]{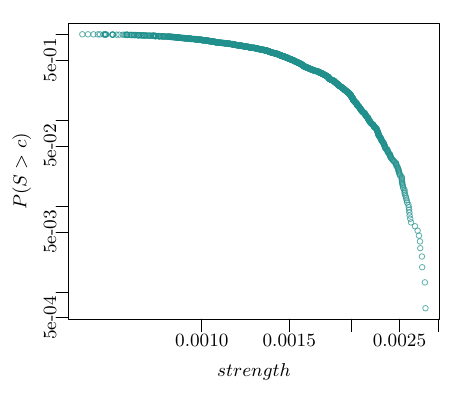}
            \includegraphics[width=\linewidth,height=3.5cm,keepaspectratio]{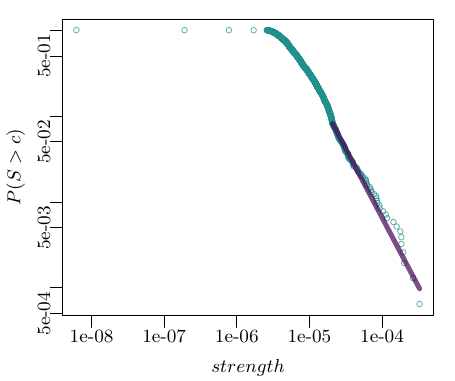}
            % \caption{Model training workflow diagram}           
        \end{minipage}
        % \hfill        
        % \begin{minipage}[t]{0.32\textwidth}
        %     \centering
        %     \includegraphics[width=\linewidth,height=3.5cm,keepaspectratio]{strengthtim48hub.pdf}
        %     % \caption{Performance metrics comparison}
        %     \label{fig:metrics}
        % \end{minipage}
        \caption{}
         \label{fig:upper}
    \end{subfigure}
    
    % Bottom: Single wider subfigure
    \vspace{0.5cm} % Vertical spacing
    \begin{subfigure}{\textwidth}
        \centering
        \includegraphics[width=1.1\linewidth,keepaspectratio]{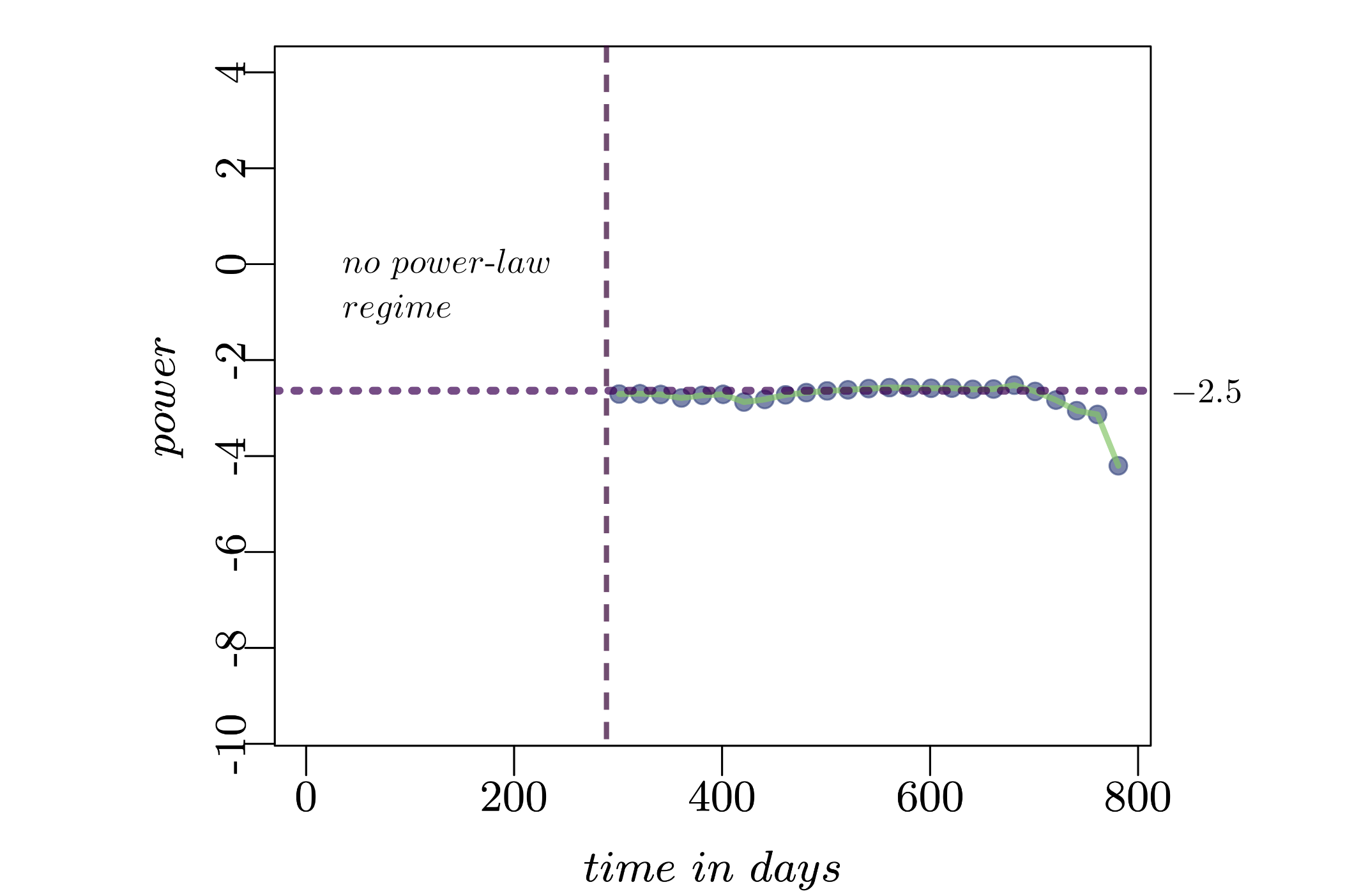}
        \caption{}
        \label{fig:lower}
    \end{subfigure}
\end{minipage}
\caption{In Fig.\ref{fig:left}, we consider a representative sample of nodes in the transportation network and follow the evolution of their strength over time, revealing the near alignment of peak strengths across the network, and this around $t_{peak} \simeq 260$ days. In Fig.\ref{fig:upper}, we show the cumulative strength distribution in the BH model for snapshots at time $t \in \{t_1=50,t_2 = 180,t_3= 360,t_4=400\}$. Distributions at $t_3$ and $t_4$ feature a distinct power-law regime with exponent $\{-2.65, -2.57\}$, %-2.55, -2.56, 
initiated around an  $s_{min}$ of %\{1.69 \times 10^{-5},1.37\times 10^{-5},
$\{8.65\times 10^{-6},4.98 \times 10^{-6}\}$ respectively. In Fig.\ref{fig:lower}, we show how the exponent of the power law experiences a transition around $t_{trans} \simeq 280$ days from no power-law regime to one with a power 
$\sim -5/2$. The transition, which is marked with a dashed line in Fig.\ref{fig:left} occurs soon after node strength reaches its global maximum across the network.  The exponents were recovered with {\sf poweRlaw}, an R-package that provides an estimate of the exponent as well as the minimum strength $s_{min}$ beyond which the power-law holds \cite{clauset2009power}.}
    \label{fig:powerBH}
\end{figure}

We probe BH-model time series, solutions of Eqs.\ref{eq:model}, for the time-evolving node strength distributions. In Fig.\ref{fig:left}, we follow the time evolution of node strength over the network which reveals a near alignment of peak strength across the network, a feature which is as intriguing as the infectivity of plateau in Fig.\ref{bhjnt}. Sectioning strength profiles at various stages in the evolution, we end up with the strength distributions of Fig. \ref{fig:upper}, which, as evident in Fig.\ref{fig:lower}, exhibit a sharp transition to a power-law-tailed regime with a rather stable power of $-5/2$, stable that is till the disease runs its course through the network.  

\begin{figure}[!htp]
\centering
 \begin{subfigure}[b]{0.46\textwidth}
        %  \centering
\includegraphics[width=1.1\textwidth]{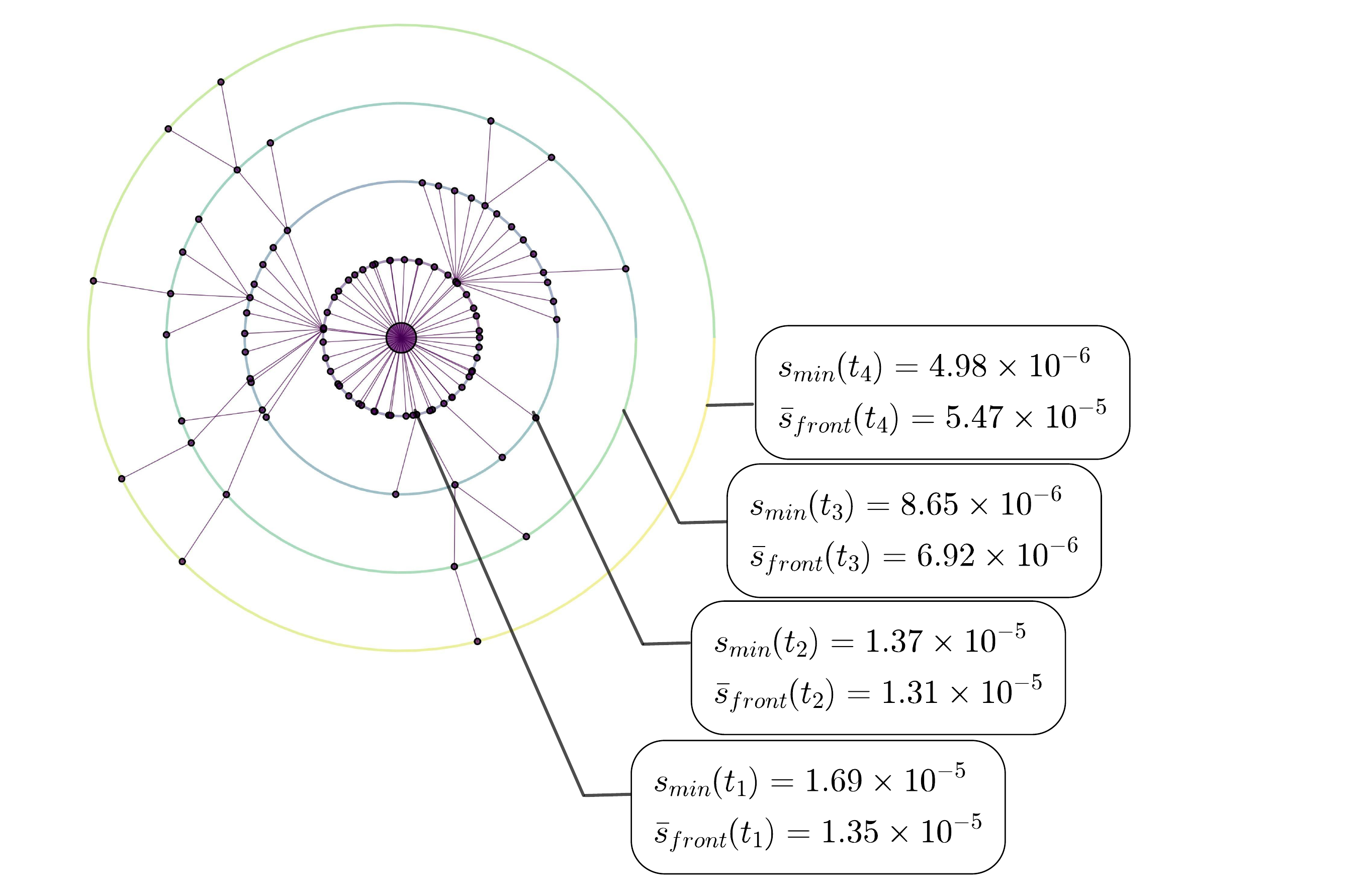}
         \caption{}
         \label{fig:powerBH2a}
     \end{subfigure}\hfill
     % \hfil
          \begin{subfigure}[b]{0.5\textwidth}
        %  \centering
    \includegraphics[width=0.9\textwidth]{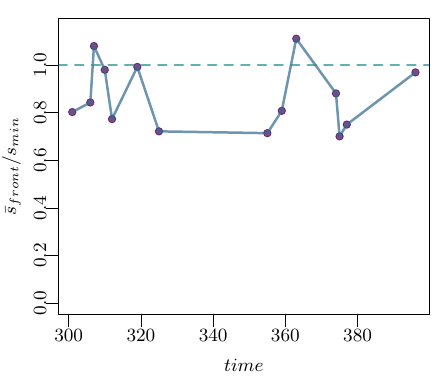}%sminandFront2%sminandFront8hubs.pdf
         \caption{}
         \label{fig:powerBH2b}
     \end{subfigure}

    %  \hfill

%               \begin{subfigure}[b]{0.475\textwidth}
%        %  \centering
%    \includegraphics[width=0.5\textwidth]{originofinfection2.pdf}
%         \caption{}
%         \label{fig:powerBH2c}
%              \end{subfigure}
%                       \begin{subfigure}[b]{0.475\textwidth}
%        %  \centering
%                  \includegraphics[width=0.7\textwidth]{ModulOriginNoSource}
%         \caption{}
%         \label{fig:powerBH2d}
%
%     \end{subfigure}
     
%\includegraphics[width=0.3\textwidth]{BHnoControlWithSource.pdf}
\caption{ We follow strengths over the network in the interval $t\in [280,400]$. The interval coincides with the emergence of an infectiousness plateau, the formation of fronts and their steady propagation, leaving invariant power-law strengths distribution in their wake. Monitoring disease arrival times, we recover BH wave fronts, identify nodes which span them, and calculate their mean strength $\bar{s}_{front}(t)$. A sample of such nodes for $t \in \{t_1=280,t_2 = 320,t_3= 360,t_4=400\}$ are shown in Fig. \ref{fig:powerBH2a}, together with $\bar{s}_{front} (t)$,  and $s_{min}(t)$  for those four times. We note that $\bar{s}_{front} (t)$ is remarkably close to  $s_{min}(t)$, a pattern which is born out at other times as evident in Fig. \ref{fig:powerBH2b}. This strengthens the conjecture that the minimum strength in the power law distribution $s_min(t)$ obtains on or close to the evolving wave front at time $t$, with the distribution of strengths organized in an invariant power law behind it. This is further correlated with the unfolding infectiousness plateau with which the reported behavior overlaps. }
\label{fig:powerBH2}       % Give a unique label
\end{figure}

When established, the $-5/2$ power-law has a minimum strength $s_{min}$ below which it fails to hold and a maximum strength beyond which statistics are no longer reliable. Both ends of the distribution are decreasing with time, with the spread of disease over the network. We suspect $s_{min}$ at any given time to be correlated with the position of the BH front at that time with the distribution self-organizing in its wake. Using the BH framework, we identify nodes at the front and compare their average strength $\bar{s}_{front}$ with $s_{min}$. We perform this procedure soon after the BH front is established, which interestingly enough coincides with the time when the transition to power-law of $-5/2$ occurs, around $280$ days, then move forward till the front hits the edge of the network, around $t=400$ days. During this phase, $\bar{s}_{front}$ with $s_{min}$ are evidently quite close, with the ratio fluctuating near unity as shown in Fig.\ref{fig:powerBH2b}. This suggests that the power law distribution trails behind the BH reaction-diffusion front, with nodes of minimum strength $s_{min}(t)$ on or close to the front at time $t$. From this perspective, the closest nodes to the outbreak location are recruited early on and their strength distribution converges to the $-5/2$ power, followed by the more distant ones with weaker strengths, which nonetheless are still governed by the same power-law as shown in Fig.  \ref{fig:powerBH}. Therefore, concomitant to the wavefront propagation, there is a recruitment of sub-graphs whose strengths are getting weaker all the while their nodes' strength distribution is self-organized in  $-5/2$ power-law, with an inverse relation between typical strength and effective distance. 

Interestingly enough, the transition to the power law regime occurs around the emergence of the plateau in $j_n$ [the inception of which was earlier correlated with the emergence of BH-fronts], and soon after node strength reaches it global maximum over the network. Growth and decay of infectivity as disease propagates down the network converges to a node-invariant pattern all the while strength distribution cascades down in mean value, while maintaining the $-5/2$ power-law structure. One is tempted to speculate that the subsiding of the initial transient, then the democratisation of infectivity and recovery across the network are the harbingers of a transition to a cascading steady state which is now controlled by the emergence of communities in the networks. We speculate but have to relegate a thorough analysis of the timing of the transition then the tight correlation between power-law and plateau to future works, as we focus on the power-law regime itself, and proceed to construct a kinetic model which captures the recovery of previously infected nodes, together with recruitment of more and more nodes in the propagating wave. What we are after is a Master equation which encodes disease specific kinetics with the aim of capturing [in a natural regime of disease propagation] the steady power law distribution of strengths shown in Fig. \ref{fig:UPP}, with a lower cut-off $s_{min}$ which is shifting to smaller values as evident in Fig. \ref{fig:powerBH2a} and \ref{fig:powerBH2b}. 

% We start by re-scaling the strengths by $s_{min}$: this will set the lower bound in strength to 1, a renormalizing step which puts us on course to the scale-invariant power law we are after. 
We start by discretizing strengths into bins with strength $S^k$, each associated with a probability $P_{S^k}$. We have defined the bin strength with a capital $S$ and a superscript $k$ to differentiate it from that of a node $m$ with strength $s_m$. A bin of strength $S^k$ will thus assemble all nodes $m$ with strength $s_m = S^k$. We renormalize the strength by $s_{min}$ so that the lowest strength $S^1 =1$.

Motivated by kinetic theory of aggregation, in which clusters of monomers or polymers pair up and merge into a larger-sized single cluster, and its extension to social contexts, we aim to write a master equation describing the evolution of strength distribution, where strength $S$ defines the infectivity export of a given node. 

We postulate that the equation should account for binary interactions, reflecting the exchange of infectivity between localities with different strengths, as well as the intrinsic infectivity evolution in localities of the same strength. 

Following the logic of aggregation dynamics, we chose the binary interaction kernel to be multiplicative in the strengths, which leads to a second order term $(S_k P_k)$. Then to reflect the intrinsic recovery we postulate that it is first order in the combination $S_k P_k$. The evolution equation is then given by: 

\begin{equation}\label{agree}
 \frac{\partial P_{S^k}}{\partial t} = - r S^k P_{S^k} - P_{S^k} S^k  \sum\limits_{i=1}^{} S^{i} P_{S^i}  + a \sum\limits_{i=1}^{i<k} S^i P_{S^i} S^j P_{S^j}  
\end{equation}

The first term appearing in Equation \ref{agree} accounts for the loss in the weights due to intrinsic recovery, with a rate $r$. The second term accounts for the loss in $P_{S^k}$ due to nodes in the $ S^k$ bin growing to a higher strength values  $S^{k'}$ and thus moving out of the $S^k$ bin. This happens though import of infectivity through a binary interaction with any other bin of strength $S^i$. The rate of this process is taken as a unit rate. 
Lastly, the third term describes the gain in $P_{S^k}$ due to mutual exchange of infectivities between nodes whose strengths $S^i$ and $S^j$ add up to $S^k$. The coefficient $a$ is the relative rate of the process.  

We note that the bin of the smallest strength  is different from the other bins in the way its weight evolves. 
More specifically,  the loss term is solely due to binary interactions with others $S^j$. The intrinsic loss is by definition absent, because $s_{min}$ is the lower boundary of the power-law region. Further, the binary gain term is absent for the same reason. Instead, we postulate that the instrinsic loss in the higher strengths bins $S^k$ brings them directly to the lowest bin. Hence, their sum constitutes the linear gain term in the evolution of the weight of the smallest bin. Its equation is given by the below: 
\begin{equation}\label{eq1}
     \frac{\partial P_{S^1}}{\partial t} = r \sum\limits_{i>1}^{\infty} S^i P_{S^i}   -  P_{S^1} \sum\limits_{i>1}^{\infty} S^i P_{S^i}
\end{equation}

It is worth noting that in contrast to cluster aggregation models with the underlying mass conservation law, no analogue is to be found in the infectivity dynamics. However, the general idea of identifying two types of dynamic drivers, namely binary interactions (exchange of infectivity strength with neighboring nodes) and intrinsic isolated node dynamics, with the corresponding second and first-order terms in the evolution equations seems well justified. 

The form of the equations \ref{agree} and \ref{eq1} that we essentially postulate has two redeeming features:  it allows a non-vanishing stationary set of statistical weights  $P(k)$ and an asymptotically exact analytical solution for this stationary state. A laborious though straightforward calculation shows that the above system admits a steady state scale-invariant solution of the form  $P_S \propto S^{-5/2}$. This is remarkably consistent with observed cascade of the strengths distribution to smaller and smaller strengths as the epidemic percolates through the network, all the while maintaining the same power-law distribution over a range of strengths that remain governed by the above kinetics. We refer the reader to Sec.\ref{appb} of the Supplementary Materials for derivation details, noting that it relies principally on the generating function approach already exploited in \cite{krapivsky2010kinetic,johnson2006universal}. 

In addition to the power law regime which is evidently universal with no dependence on model parameters, the steady state limit features an exponential which depends on the coupling constant $a$, and which is decaying except at the critical value $a=3/4$. Note that the recovery rate $r$ does not affect the shape of the steady state distribution: this is perhaps not surprising for while $r$ contributes the rate at which the distribution evolves to the steady state, and may affect the overall amplitude (as it does), it is not expected to manifest in the balancing act which controls the resulting distribution.

As with other instances of exactly solved models the insights brought by the analytical solution turn out to be of more general importance. By numerically exploring a broader class of evolution rules we demonstrate that they all belong to the same universality class. Within this class an appropriately defined stationary distribution is characterized by the common $(-5/2)$ power-law modified by an exponential term; the exponential correction disappears under critical parameter conditions and remains insignificant in a near-critical domain \footnote{We note that Equation \ref{agree} governs the forward dynamics of the wave propagation from the source to the extremities in concentric circles, where $s_{min}$ and $\bar{s}_{front}$ are shown to be comparable. The power law still obtains over a temporal domain beyond that defined by the longest effective distance $D_{eff}$ of BH. We expect that the back-propagation of the wave from the extremities back to the center to explain its maintenance where the effects appearing in Equation \ref{agree} are still present and switch magnitudes.}

We have shown that BH reaction-diffusion dynamics cascades over a network as it leaves behind a wake of infected nodes distributed in a scale-invariant power law which we then proceeded to recover in a kinetic model which is grounded in disease propagation between localities in the network. The success of SIR driven BH models would seem to suggest that this numerical cum theoretical conclusion will quite likely be born out by careful analysis of surveillance data in an emergent epidemic. As it happens, the very recent global COVID-19 pandemic gave us all plenty to ponder as it diffused reactively, infectively, over the planet. Leaving various critical caveats to the discussion, we proceed to further substantiate our theoretical findings with COVID-19 evidence. 

We first point the reader to our recent results with stochastic modeling of the evolution of the daily-locality-level infections in Lebanon \cite{najemtouma1}, which leverages the framework developed in \cite{hhh,meyer2014power,meyer2014spatio,ssentongo2021pan}. Here one works with an auto-regressive model out of which we were able to abstract a time evolving network which maps naturally to the one we associated with BH dynamics. In particular, the distribution of ``strengths" in that network was shown to converge, following another sharp transition, to a steady state distribution with a power of $-5/2$!

Reassuring for sure, but we had concerns about the quality of the data, given the limited resources of a country in financial collapse \cite{zahreddine2022challenges}. So we went for a more ambitious exercise, taking on publicly available US COVID-19 data. Following the procedure in \cite{irons2021estimating}, we recovered the state-level $j(t)$ of COVID-19 cases in the US from daily infection and death counts.  In doing so, and given that reporting is marred with delays and biases, we adopted a Bayesian framework to get reliable estimates of population prevalence. The framework assumes a discrete-time SIR model of the disease which recovers the time-varying reproductive number $R(t)$ through Bayesian inference. Its likelihood depends on confirmed cases, deaths, and the number of tests. Given the inferred state-specific time-dependent $R(t)$ we could estimate the value of $j_m(t)$ for all states $m$. Then using data on traffic flows $P_{mn}$ between states \cite{kang2020multiscale}, we were able to reconstruct the state level $A(t)$ for the first 16 months of the start of the pandemic. The power of the resulting strength distribution is shown below in Fig. \ref{counties}. We note that there is a transition to $-2.5$ near the fourth month of the pandemic,  which is what we expect the power to converge in the long time limit or steady state. Given the uncertainties, the different levels of preparedness, the availability of testing, and the different control strategies across the states, we expect the power of $-2.5$ to emerge at a later stage in the timeline of the disease (if at all).

\begin{figure}[!htp]
\centering
\includegraphics[width=0.4\textwidth]{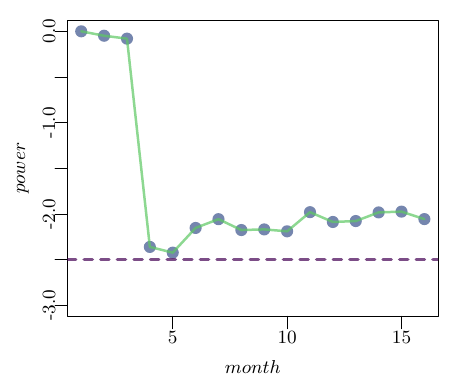}
\caption{Relying on the number of reported cases, deaths, and the number of tests in the US, we recovered the infection counts $j(t)$ at the state level. For this, we used a Bayesian framework to infer the state-specific time-dependent $R(t)$,  which allows us to estimate $j(t)$.  We then combine $j(t)$ over states with the interstate traffic $P$, which is only available for the 16 first months of the disease, in order to reconstruct the temporally evolving matrix $A = j P$.  The figure shows the evolution of the recovered power of the state-level strength distribution of $A$.  We note that it exhibits a transition around the fourth month of the pandemic to the value of -5/2, represented by the dashed purple line, and fluctuates thereafter around -2.1.}
\label{counties}       % Give a unique label
\end{figure}

We have presented three complementary perspectives on the emergence of scale-invariant power law distributions of infection strength as a disease cascades over a generic transportation network. The first shows it emerging from dynamical simulations of node-based infections transported across the network. The second recovers it as a scale-invariant solution of a Master equation encoding the kinetics of disease spread and recovery across the network. The third detects the same transition to a scale-invariant regime in stochastic modeling of surveillance data. Taken together, our results speak for the universality of such a transition, and the power law structure that emerges in the wake of reaction-diffusion evolution to localities in a sufficiently developed network. The detection of this regime in surveillance and transportation data can be taken as a measure of unfettered systemic, infectivity across the network. To be sure, the implied universality of this regime begs for further testing over networks with highly resolved surveillance and transportation data. Further substantiation with sophisticated dynamical models of networked infectivity will then put it at par with Kolmogorov-Obukhov theory of turbulence.

The reported scale invariant power-law regime reflects a time-evolving cascade which must of course terminate as an emergent disease plays itself out over susceptible populations in a finite network. But even before then, control measures will work to disrupt the free-flow of infectivity over the transportation network (which is taken to be time-invariant in our theory). The absence of control, or free-flow, is reflected in the constancy of $r$ in the Master equation. It is worth noting that choosing $r$ to be strength dependent, $r \rightarrow r/s^{\alpha}$ in  Eq. \ref{fissionfusion} leads to a generalized $\alpha-$dependent steady state $P_S \propto S^{-(2.5 - \alpha)}$. 

In the course of our multilayered exploration, we highlighted the emergence of a node-invariant pattern of growth then decay in infection counts, then further noted how: i- its emergence correlates neatly with the establishment of BH-fronts, then the transition to power-law distributions; ii- its unfolding overlaps with, as it extends BH-front propagation, and spans the bulk of the power-law cascade of the disease. Revisiting the plateau of Fig.\ref{bhjnt} with scale-invariant strengths distributions in mind, we asked whether network-wide distributions in infection counts self-organize in a time-invariant power-law that somehow reflect the node-invariant pattern of growth and decay? As briefly reported in Sec.\ref{j-dist} of the Supplementary Material, the answer is yes, eventually, i.e after the epidemic has permeated the network, and till it plays itself out. This , together with the correlations noted above, are begging for a rigorous explanation, partly addressing the self-organization of SIR dynamics over networks, but also translating strength kinetics into that of infectiousness.  

Last but not least, we note that the reported power-law in strength distribution was previously observed then modeled in the evolution of insurgency across a wide variety of conflicts; then further detected in the herding behavior of the financial market, and more generally in binary aggregation/fragmentation phenomena \cite{johnson2006universal,johnson2011equivalent,ruszczycki2009relating,bohorquez2009common,eguiluz2000transmission,krapivsky2010kinetic} which of course underlie the processes driving disease kinetics.  It is thus compelling to think of disease spread and recovery in relation to the processes of coalescence and fragmentation in warring groups, or perhaps more more productively, to approach the "statistics of deadly quarrels" as one would approach that of epidemics, then perhaps put ourselves in a better position to capture the notoriously intimate coupling between wars and diseases.

% Your references go at the end of the main text, and before the
% figures.  For this document we've used BibTeX, the .bib file
% scibib.bib, and the .bst file Science.bst.  The package scicite.sty
% was included to format the reference numbers according to *Science*
% style.

%BibTeX users: After compilation, comment out the following two lines and paste in
% the generated .bbl file. 

\bibliography{scibib}

\bibliographystyle{Science}

\section*{Acknowledgments}
We acknowledge the support of the Ministry of Public Health and the Center for Remote Sensing in Lebanon for data access which led us to the observations upon which our work in this manuscrupt is based. 
\subsection*{Funding}
The authors acknowledge that they received no funding in support for this research.
\subsection*{Authors contributions}

\subsection*{Competing interests}
The authors declare that they have no competing interests.
\subsection*{Data and materials availability} The codes used in this paper are available on the following link.
%Here you should list the contents of your Supplementary Materials -- below is an example. 
%You should include a list of Supplementary figures, Tables, and any references that appear only in the SM. 
%Note that the reference numbering continues from the main text to the SM.
% In the example below, Refs. 4-10 were cited only in the SM.     
\section*{Supplementary materials}
Materials and Methods\\
Supplementary Text\\
Figs. S1 to S3\\
%References \textit{(4-10)}
\section*{Material and Methods}
\section{Model Solution and Associated Methodology}\label{sec1app}
\begin{itemize}
\item We calculate the nodes' betweenness centralities to identify the hubs, which are the top eight localities. The rest of the localities have significantly lower centrality (less the $10\%$ compared to the one with the largest centrality). We initialize their corresponding $j_n(0)$ to a random value weighted by their populations. The recovered population $r_n$ is set to 0 and susceptible $s_n$ is calculated following $s_n +j_n + r_n = 1$
\item We then solve Eq. \ref{eq:model} given by:
\begin{eqnarray*}
\partial_{t}j_{n} & = &a v_{n}j_{n}-b j_{n}+c\sum_{m\neq n}^NP_{mn}\left(j_{m}-j_{n}\right),\nonumber \\
\partial_{t}v_{n} & = &-a v_{n}j_{n}+c \sum_{m\neq n}^N P_{mn}\left(v_{m}-v_{n}\right),
%\label{eq:model}
\end{eqnarray*}
with this choice of parameters: $b= 0.05$, $a= 1.5b$, $c = 1e^{-6}$,  $N = 1544$, and the above described initial conditions. 
\item We choose the flow between localities $P_{mn}$ to be equal to $I_m F_n D_{euc,mn}^{-2.6}/\sum P_{mn}$, with the Euclidean pairwise distances $D_{euc,mn}$ calculated between the centroids of Lebanon's localities, and the populations $I_m$ and $F_n$ taken from \cite{najem2022framework}. We use another definition of distance in which the latter is replaced with adjacency orders $O_{ij}$ between localities, which is used in \cite{meyer2014power}. For this purpose, we read Lebanon's shapefile as an {\sf SpatialPolygons} object
of the {\sf sp} package to derive the matrix
of adjacency orders automatically using the functions {\sf poly2adjmat} and {\sf nbOrder} of the {\sf surveillance} package in R . Given an adjacency matrix, the function {\sf nbOrder} determines the integer matrix of
shortest-path distance.  All of the below results remain robust to this variation. 
\item We define the matrix $A(t) = j(t)P$ and follow its nodes' strengths defined to be $s_n = \sum j_n P_{nm}$
\item We feed the strengths vector to {\sf conpl}, {\sf conlnorm}, and {\sf conexp} functions which correspond to 
continuous power-law,
continuous log-normal, and
continuous exponential of the {\sf poweRlaw} package in R in order to characterize the distribution type. The choice of the functions is justified by the fact that the strength can take on any positive value and is not discrete. In the latter case the corresponding functions are used: {\sf displ} (discrete power-law),
{\sf dislnorm} (discrete log-normal), and {\sf disexp} (discrete exponential)
\item We compare the goodness of fit between the different distribution types using {\sf compare\_distributions} of the {\sf poweRlaw}, which is a likelihood ratio
test for model selection using the Kullback-Leibler criteria.
\item We report that the power-law is the best distribution describing the data for each time frame. Thus we keep track of its exponent and $s_{min}$,  which is the minimum strength below which the power law fails to hold using {\sf estimate\_xmin} of the {\sf poweRlaw} package. These are reported in Figures \ref{fig:powerlawregime} and \ref{fig:powerBH}.
\item We feed the time series $j_n(t)$ to {\sf NetOrigin} to recover the source of infection. It returns the identity of the source using the effective distance median (edm), in which mean effective distance  and standard deviation are computed from each potential origin of infection, and the node that simultaneously minimizes both these quantities is identified as the source since their minimization is associated with increased concentricity.

\item Having identified the origin of infection we calculate the effective distance between the source and all the other localities, which is defined to be $(1-\log{P_{mn}})$.

\item We find each node's arrival time, which is defined to be the first time the product of the locality's $j_n$  with its corresponding population reaches the interval $[1,2)$. Evidently, the ratios of effective distance to the arrival times are expected to be constant, which physically denotes the speed of propagation of the wave. For these arrival times $t \in \{t_1=280,t_2 = 320,t_3= 360,t_4=400\}$, the respective effective distances are given by $D_{eff} = \{10.96, 11.10, 11.31, 11.39\}$. We note that the nodes at $D_{eff}  = 11.39$ correspond to the following localities Jouaya [88km], Babliye [64km], Bebnine [102km], Qnaiouer [108km] from Beirut.
\item We identify the nodes associated with each arrival time and record their mean strength, which we denote by $\bar{s}_{front}$. These are the nodes partaking in the front.
\item We finally follow the dependence of the $\bar{s}_{front}$ on $s_{min}$ and their ratio goes to 1 at the time the power law with exponent -2.5 is established, which is reported in Figures \ref{fig:powerBH2a} and \ref{fig:powerBH2b}.
\item This exercise was carried over with different initial conditions of infectivity in Lebanon's localities and proved to be robust to variations.  We also tried our framework on the  Austria's network and recovered all the features found in our simulations of BH for Lebanon. 

\end{itemize}
\section{Characterization of the $j(t)$'s distribution}\label{j-dist}

\begin{figure}[!htp]
\centering
 \begin{subfigure}[b]{0.5\textwidth}
        %  \centering
\includegraphics[width=0.9\textwidth]{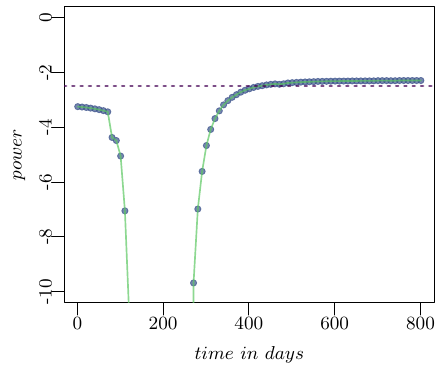}
         \caption{}
         \label{fig:powerj}
     \end{subfigure}\hfill
     % \hfil
          \begin{subfigure}[b]{0.5\textwidth}
        %  \centering
    \includegraphics[width=0.9\textwidth]{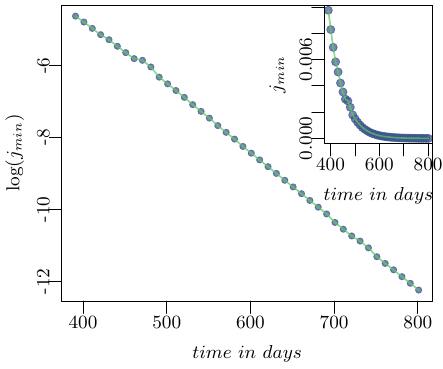}%sminandFront2%sminandFront8hubs.pdf
         \caption{}
         \label{fig:jmin}
     \end{subfigure}

\caption{ The power law the distribution of $j(t)$ is followed in Figure \ref{fig:powerj}.  It transitions to -2.5 at around $t=400$.  The transition in the distribution of strength to the same power precedes that of $j$ by approximately 120  days. Also, the evolution of $j_{min}(t)$ is followed in Figure \ref{fig:jmin} on a semi-logarithmic plot with the inset showing the exponential decay of $j_{min}$ with respect to time. }
\label{fig:powerjjmin}       % Give a unique label
\end{figure}

 The evolution of $j(t)$ reported in Figure \ref{bhjnt} made us wonder about the significance of the plateau in relation to the emergence of power-law strength distribution, and whether there is an equal self-organization processes happening at the level of $j(t)$ as well. Therefore, we follow the distributions of $j(t)$ for the course of the evolution of the disease and also note that they converge to a power-law with an exponent -2.5 with log-linearly decaying lower cutoffs $j_{min}$, as shown in Figures \ref{fig:powerj} and \ref{fig:jmin}. It is worth noting though that this convergence to -2.5 is preceded by the convergence of the strength distributions to that same power with a lead time of about 120 days.  
The delay hints that $P_{mn}$ contributes to the emergence of the power-law in the distribution of $s$, for about 120 days before $j$ takes over and together with $P_{mn}$ lead to the steady behavior of the $s$ distribution. 

\clearpage  

\section{Solution of the Master Equation}\label{appb}

Here we report on the details behind the steady state solution of the equations governing the kinetics of strength dynamics which were proposed in the body of the text: 

\begin{subequations}\label{fissionfusion}
  \begin{empheq}[left=\empheqlbrace]{align}
    \frac{\partial P_{S^1}}{\partial t} &= r \sum\limits_{i>1}^{\infty} S^i P_{S^i}   -  P_{S^1} \sum\limits_{i>1}^{\infty} S^i P_{S^i} \\
    \frac{\partial P_{S^k}}{\partial t} &= -rS^kP_{S^k} -   S^kP_{S^k} \sum\limits_{i=1}^{\infty} S^i P_{S^i}  + a{ } \sum\limits_{i=1}^{i<k} S^i P_{S^i} (S^k-S^i)P_{S^k-S^i}. 
     \end{empheq}
\end{subequations}

We start with an attempt towards the asymptotic distributions which makes use of generating functions. We complement this treatment with the details of the numerical explorations which confirm as they extend our analytical results. 
\subsection{Analytical Solution}
We focus on the properties of the steady state defined by the evolution Equations \ref{fissionfusion} (using the generating function method). Importantly, the equations do not support a conservation law of the form  $d \sum P(S^k )/dt=0$  and hence, the variables $P(S^k)$ should be interpreted as statistical weights rather than probabilities proper.  This does not contradict the original context of node infectivity strength, where no conservation law is expected a priori. 
%Whether a steady-state collection of weights exists at all in the absence of the conservation law is an open question and we address it in the context of exploring modifications of the evolution equations.

% Our aim here is to characterize the steady-state solution of the master Equation \ref{fissionfusion} , and to confirm that it indeed follows a power-law with an exponent of $=-5/2$. 
For this purpose, we start with the $P_{S^1}$ equation, noting that the procedure carries over to $S^k>1$. We start with the steady state of $P_{S^1}$, that is $\frac{\partial P_{S^1}}{\partial t} = 0$, which is given by the below:\\

\noindent
\begin{equation}\label{ssof1}
  % \frac{\partial P_{S^1}}{\partial t} = r \sum\limits_{i>1}^{\infty} S^i P_{S^i}   -  P_{S^1} \sum\limits_{i=1}^{\infty} S^i P_{S^i} 
  % \\
  % \hfil
  r \sum\limits_{i>1}^{\infty} S^i P_{S^i}  - P_{S^1} (\sum\limits_{i>1}^{\infty} S^i P_{S^i} ) = 0 
   \end{equation}
\noindent
We introduce the generating function $G[y]$, with the summation starting at index $i=1$. The latter is then split into the term corresponding to the $i=1$ term of the series and a summation that starts with $i>1$ for reasons that will become clear as we proceed. It is given by: 
\begin{equation}\label{gnf}
G[y] = \sum\limits_{i=1}^{\infty} S^i P_{S^i}  y^{S^i} = P_{S^1} y + \sum\limits_{i>1}^{\infty} S^i P_{S^i}  y^{S^i} = P_{S^1} y  + g[y],
 \end{equation}
 \noindent
 where we define the last term of the above equation as:
 \begin{equation}\label{lgnf}
g[y] = \sum\limits_{i>1}^{\infty}  S^i P_{S^i} y^{S^i}
 \end{equation}
Evaluating Equation \ref{lgnf} at $y=1$ we can write Equation \ref{ssof1} as: 

\begin{equation}\label{systemeq1}
r g[1] - P_{S^1} g[1] =0 
 \end{equation}
 which gives: 
 \begin{equation}\label{p1ana}
P_{S^1}  = r
 \end{equation}

 \noindent
Using the definition of the generating function of Equation \ref{gnf} in Equation \ref{fissionfusion}(b) we get: \\

\noindent
 $ -rS^jP_{S^j} + a  \sum\limits_{i=1}^{\ {i<j}} S^i P_{S^i} (S^j-S^i)P_{S^j-S^i} - S^jP_{S^j}  {\sum\limits_{i=1}^{\infty} S^i P_{S^i}} = 0 $
\\
 $ rS^jP_{S^j}  +  S^jP_{S^j}    {G[1]}=  {a }\sum\limits_{i=1}^{{\infty}} S^i P_{S^i} (S^j-S^i)P_{S^j-S^i} $\\
 $ S^jP_{S^j} (r +   {G[1]})  =  a{} \sum\limits_{i=1}^{\infty}S^i P_{S^i} (S^j-S^i)P_{S^j-S^i} $
 \\
 
 \noindent
 The above now can be written as: 
\begin{equation}\label{ns1}
S^jP_{S^j}   =  \frac{a}{(r   +   {G[1]})}\sum\limits_{i=1}^{\infty}S^i P_{S^i} (S^j-S^i)P_{S^j-S^i}, 
 \end{equation}
 
 \noindent
Summing over the $j>1$ in Equation \ref{ns1} and using the definition of $g[y]$ given in Equation \ref{lgnf} we get:
 \\
 
 \noindent
  $ \sum\limits_{j>1}  P_{S^j} S^j y^{S^j} = g[y]  =\frac{a}{(r   +   {G[1]})}  \sum\limits_{S^i} \sum\limits_{S^i}   P_{S^i} S^i (S^j-S^i) P_{S^j-S^i}y^{S^j}$
  \\

  \noindent
 The right-hand side of the above equation involves a summation over two dependent indices through $S^j - S^i = S^k$, which when substituted for in the above equation gives:
  \\

  \noindent
   $ g[y] = \frac{a }{(r  +    {G[1]})} \sum\limits_{S^k} \sum\limits_{S^j}   P_{S^j} S^j (S^k) P_{S^k}y^{S^j + S^k}$\\
      $ g[y] =  \frac{a }{(r  +    {G[1]})} \sum\limits_{S^k} \sum\limits_{S^j}   P_{S^j} S^j y^{S^j } S^kP_{S^k}y^{S^k}$\\
        $ g[y] = \frac{a }{(r   +   {G[1]})} {\sum\limits_{S^k}} { \sum\limits_{S^j}   P_{S^j} S^j y^{S^j }}{S^k P_{S^k}y^{S^k}}$\\
        
\noindent
Each of the summations is carried out over a dummy index starting from 1, which we identify as $G[y]$, and thus the product of the series is nothing but $G[y]^2$. This gives us an equation of $g[y]$ as a function of $G[y]$:
 
 \begin{equation}\label{generalgy}  g[y] = \frac{a }{(r   +    {(P_{S^1} + g[1])})}  G[y]^2 
  \end{equation}
 
 \noindent
 Evaluating Equation \ref{generalgy} at $y=1$ in order to couple it with Equation \ref{systemeq1} we get:
  
   \begin{equation}\label{generalg1}  g[1] = \frac{a }{(r   +    {(P_{S^1} + g[1])})}  G[1]^2 =   \frac{a }{(r   +    {(P_{S^1} + g[1])})}  (P_{S^1} + g[1])^2 
  \end{equation}
% wolfram alpha  
%  solve (1 - m) x^2 + 2 m y - (1 - m) y^2 = 0
%(1 - m) x^2 - m y + (1 - m) x y = 0

\noindent
  Equations \ref{generalg1} and \ref{p1ana} give the system of equations: 
  \begin{empheq}[left=\empheqlbrace]{equation}
    \begin{aligned}
     & P_{S^1} = r \\
     & - rg[1] + a G[1]^2 -  g[1]G[1] = 0
    \end{aligned}
\end{empheq}
\\

% Wolfram Alpha
%solve (1 - m) x^2 - m y + 4 (1 - m) x y + 3 (1 - m) y^2 = 0
%-2 (1 - m) x^2 + m y - 2 (1 - m) x y = 0
\noindent
The latter gives a quadratic equation in $g[1]$ whose solutions are:\\

\noindent
$g[1] = - r \pm r \sqrt{\frac{1}{1-a}}$ \\

\noindent
These together give: 
\\

\noindent
$G[1] = g[1] + P_{S^1} = \pm r \sqrt{\frac{1}{1-a}} $

\noindent
 When $G[1] =  r\sqrt{\frac{1}{1-a}}$ is plugged into the denominator of Equation \ref{generalgy} it gives a quadratic equation in $g[y]$. We define $c = \sqrt{\frac{1}{1-a}}$.

 \begin{equation}\label{generalgyupdated}  g[y] = \frac{a }{(r   +    G[1])}  G[y]^2 =  \frac{a }{r(1   +c)}  (  r y + g[y])^2 
  \end{equation}
\noindent
Equation \ref{generalgyupdated} gives a quadratic equation in $g[y]$ whose solution is given by: 
\begin{equation}\label{solu0}
g[y] = \frac{-(2P_{S^1}  - B) \pm B \sqrt{1- \frac{4P_{S^1}y}{B}}}{2}
  \end{equation}
\noindent  
where $B = \frac{P_{S^1}(1+c)}{a}$.
\\

\noindent  
Using the below definition of Equation \ref{expan} together with \ref{solu0} and noting that $x = \frac{4P_{S^1}y}{B}$ we get: \\
  \begin{equation}\label{expan} 
(1-x)^{1/2} = 1 -x/2 - \sum \frac{(2k-3)!!}{(2k)!!} x^k
  \end{equation}
 \noindent 
    \begin{equation}\label{expan1}g[y] =  \frac{1}{2}[-(2P_{S^1}  - B) - B ( 1 - \frac{1}{2}\frac{4P_{S^1}}{B}    - \sum_{k>1} \frac{(2k-3)!!}{(2k)!!} (\frac{4P_{S^1}}{B}y)^k)] = \frac{B}{2}\sum_{k>1} \frac{(2k-3)!!}{(2k)!!} (\frac{4ay}{(1+c)})^k 
    \end{equation}
 where the double factorial denotes: $k!! = n(n-2)(n-4)...$\\
 
 \noindent
$g[y]$ is given in Equation \ref{lgnf} which is written as powers of $y$ as such:  
   \begin{equation}\label{expan2}
g[y]  = \sum\limits_{k>1}  k P_{k} y^{k}
 \end{equation}
Equation \ref{expan1} and \ref{expan2} give:\\ 

 \begin{equation}\label{expan4}
P_{k}  \propto \frac{(2k - 2)!}{(k!)^2} (\frac{4a}{(1+c)})^{k}
  \end{equation}
\noindent  
Finally, applying Stirling's approximation given on Equation \ref{expan4} gives: 
 \begin{equation}\label{stirling}
\ln{(k!)} = \frac{1}{2}\ln{2 \pi} + (k+\frac{1}{2}) \ln{k} - k ...
  \end{equation}
%  \begin{equation}\label{stirling}
% \ln{(S^i!)} =  S^i \ln{S^i} - S^i ...
%   \end{equation}

  \begin{equation}\label{expan5}
P_{k} = P_{1}[\frac{4a}{(1+c)}]^{k}k^{-5/2} = r (  \frac{4a}{1+1/\sqrt{1-a}})^k k^{-5/2}
 \end{equation}
\noindent

The above can now be written as: 

  \begin{equation}\label{expan6}
P_{k} = P_{1}e^{-f(a)k}k^{-5/2} = r e^{-f(a)k}k^{-5/2}
 \end{equation}
\noindent
where $f(a) = \log{\frac{1+1/\sqrt{1-a}}{4a}}$.

% Which gives: 
%  \begin{equation}\label{expan5}
% P_{S^i} = \frac{1}{2^{ 3/2}} \frac{4r}{ } {3}^{S^i} (S^i)^{(-3/2 - 1)} \approx \frac{1}{2^{ 3/2}} \frac{4r}{ } {3}^{S^i} (S^i)^{-5/2}% \frac{4r}{ }  ({3})^{S^i} 2^{2S^i -3/2} (S^i)^{(-5/2)}
%  \end{equation}
Equation \ref{expan6} recovers the expected behavior of the strength distribution, which is predominated by the power-law $P(k) \propto k^{-5/2}$, for a range bounded by $k_{max}$ defined by the maximum strength the system can exhibit given than $j_{n,max}$ is finite.
We note that the power law is universal and does not depend on the values of the model parameters. The exponent coefficient defines the rate of the exponential decay and depends only on the coupling constant $a$. It is a non-negative function of $a$ with a minimum at $a=3/4$ where the exponential term in the distribution disappears: $f(3/4)=0$. Thus, the exponentially decaying term is present except for the critical value $a=3/4$. 
As to he recovery rate, $r$, it appears only in the common prefactor of Eq \ref{expan6} and does not affect the shape of the distribution
% Below we show the results of an agent-based simulation compared to our analytical results of Equation \ref{expan5} for fixed value of $a= 0.75$ and $r= 0.4$. 

%The form of the equations \ref{expan6} that we essentially postulate has two redeeming features:  it allows a non-vanishing stationary set of statistical weights  $P(k)$ and an asymptotically exact analytical solution for this stationary state. As with other instances of exactly solved models the insights brought by the analytical solution turn out to be of more general importance. By numerically exploring a broader class of evolution rules we demonstrate that they all belong to the same universality class. Within this class an appropriately defined stationary distribution is characterized by the common $(-5/2)$ power-law modified by an exponential term; the exponential correction disappears under critical parameter conditions and remains insignificant in a near-critical domain. 

\begin{figure}[!htp]
\includegraphics[width=1\textwidth]{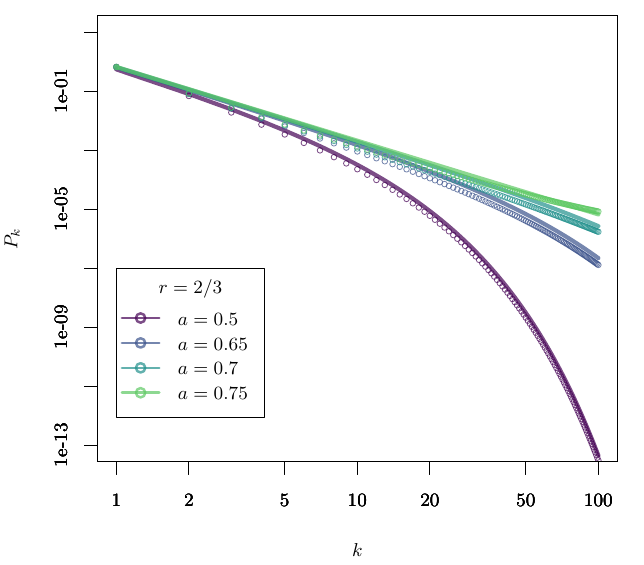}%Distributionr04
\caption{ The analytical expression of the distributions along with the steady state solutions of the model are shown for $r = 2/3$ and $a = [0.5, 0.65,0.7,0.75]$.}
\label{fig:distrAnalyticAgent}       % Give a unique label
\end{figure}

\begin{figure}[!htp]
\includegraphics[width=1\textwidth]{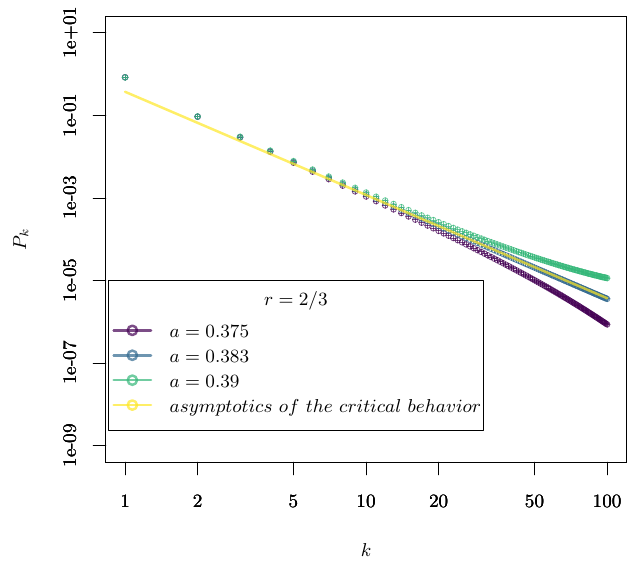}
\caption{ The analytical expression of the distributions along with the steady state solutions the equation are shown for $4 = 2/3$ and $a = [0.375, 0.383, 0.39]$.}
\label{fig:distaAnalyticAgent}       % Give a unique label
\end{figure}
\subsection{Numerical Exploration}
We verified the analytical theory predictions by numerically integrating the master equation using a simple Euler scheme. In the numerical study, the node strengths took integer values $S_k=k$  with  $1\le k \le N$ where the minimum value (after renormalization) was always taken as  $s_{min}=1$  while the maximum value $s_{max}=N$  was varied in the range   $50 \le N \le 400$. The evolution towards the steady state is robust provided the integration step is small enough,  $\le 10^{-3}$. Integration was stopped when the all the right hand side terms dropped in absolute values below $10^{-6}$, with the steady state distribution quite insensitive to the exact cutoff value. Note that the typical range,  $s_{max}/s_{min}$ ,  of the power-law distribution in the node strength as observed both in the real infection data \cite{najem2022framework} and in the networked SIR models (see Figure \ref{fig:powerlawregime}) is of order $10^2$ or less. 
The numerical results are summarized as follows:
\begin{enumerate}
\item The overall prefactor  $r$, the universal power law, $S_k^{-5/2}$, and the exponential factor as predicted by Equation \ref{expan6} are all consistent with the numerical asymptotic shape of the distribution.  As long as the values of coupling parameter, $a$, are in the range $0<a<3/4$ corresponding to the decreasing branch of the $f(a)$ function Equation \ref{expan6} is also confirmed. The steady state distribution shape is not affected by the maximum node strength $N$ (apart from the trivial effect on the range). 
\item For the coupling parameter $a$ above the critical value $3/4$  our numerical results are at odds with the theory. Instead of the decaying exponential factor as predicted by Equation \ref{expan6} we observe an increasing  exponential factor which turns out to be an essentially finite-size effect.  Namely, $f(a)\approx -\log{(N)}/N$ and vanishes in the $N \rightarrow \infty$ limit where the analytics properly applies. Note that the analytical derivation at some point employs a Taylor expansion in powers of $a$, thus assuming $a << 1$, so that the final formula can be construed as an analytical continuation to larger values of $a$. This analytical continuation fails at and beyond the critical value $a=3/4$. 
\end{enumerate}

\subsection{A broader universality class of evolution rules: Modification of the decay rate in the first-order term. }
Instead of postulating the decay rate proportional to the node strength  $rS^k$  we allow it to be strength-independent which may be naturally justified in the context of infection dynamics in an isolated locality. The changes appear in the loss term for $P(S^k )$and in the corresponding gain term for $P(S^1)$.
The assumption of strength-independent decay rate leads to a disappearance of the steady state: the right hand side vector admits only the trivial zero solution, which is not inconsistent with the terminal behavior of the full BH model. Recall, however, that to capture the intermediate self-similar regime generated in the wake of the BH propagating front we re-normalized both the smallest node strength and the statistical weights themselves. Mimicking this process, we augment the evolution of the weights $P(S^k)$ as defined by the modified Equation \ref{fissionfusion} by a normalization imposed at each integration step.
The new evolution rules do obtain a steady state. It is clear that in this case, the right hand side vector of the master equation is not zero; rather, the master equation evolution increment is canceled by the normalization step. Clearly, this implies a self-similar evolution regime preserving the shape of the distribution: each master equation evolution step amounts to multiplying the steady-state distribution by a factor less than 1, which is remedied by the subsequent normalization.
The steady-state distributions obtained numerically share all the qualitative features observed in  Figure \ref{fig:distrAnalyticAgent}. Namely, the distribution has the (-5/2) power-law factor modified by an exponential term. There is a critical line in the $(r, a)$ parameter space  where the exponent vanishes. The line separates the parameter region with low (sub-critical) pairwise coupling $a$, characterized by a dying exponential factor, from the high-a region with a weakly increasing exponential factor representing a finite-size effect.  Figure \ref{fig:distaAnalyticAgent} displays examples of stationary distributions for three parameter sets $(r=2/3, a= 0.375, 0.383, 0.39$  bottom to top)

Step-by-step normalization procedure can be also applied to the original set of master equation \ref{fissionfusion}. The steady-state distributions are quantitatively different from the non-normalized steady state solutions discussed above in the sense that the critical value of $a$ is now a function of the $r$ parameter, but otherwise demonstrate the same qualitative behavior. 
We have examined yet another generalization of the isolated bin decay rate to be proportional to some power of the node strength of the form $r(S^k)^{\alpha}$. Numerical experiments with $\infty < \alpha \le 2$ show that the evolution rules (including the step-by-step normalization) belong to the same universality class as far as the steady-state distributions are concerned. Higher values of the exponent $\alpha$ result in poor numerical convergence and have not been explored in detail.

% \section{Agent-based Simulation}
%  \label{appc}
 
%  Running an agent-based simulation with the above-described coalescence/fragmentation rules allowed us to recover the $-5/2$ power at steady-state as shown in Figure \ref{fig:agentbased}. In the simulation, a network of size $N$ is generated. The strengths of its nodes are then computed along with their distribution. A strength $S^k$ is then chosen with a weighted probability $P(S^k)$ and with a rate $r$ it transitions to 1, otherwise, another strength  $S^i$ is chosen with a weight $P(S^i)$ and with a rate $1-r$ they combine and make a strength $S^k + S^i$. The distribution is then computed and the procedure is iterated over. 
%     \renewcommand{\thefigure}{S2}

% \begin{figure}[!htp]
% \includegraphics[width=1\textwidth]{AgentBasedFissionFusion}
% \caption{The figure shows the cumulative strength distributions of networks of size $N$ for multiple realizations of the coalescence/fragmentation  $r = 0.1, 0.4, 0.8$. The respective recovered powers are $-2.54$, $-2.51$, and $-2.57$.  }
% \label{fig:agentbased}       % Give a unique label
% \end{figure}
\clearpage
\end{document}